\documentclass[10pt]{elsart}

\usepackage{amssymb}
\usepackage{graphicx}

\begin{document}

\begin{frontmatter}



\title{Transferring a symbolic polynomial expression from \emph{Mathematica} to \emph{Matlab}}


\author{A. Bret}
\ead{antoineclaude.bret@uclm.es}

\address{ETSI Industriales, Universidad Castilla-La Mancha, 13071 Ciudad Real, Spain}

\begin{abstract}
A \emph{Mathematica} Notebook is presented which allows for the transfer or any kind of polynomial expression to \emph{Matlab}. The output is formatted in such a way that \emph{Matlab} routines such as ``Root'' can be readily implemented. Once the Notebook has been executed, only one copy-paste operation in necessary.
\end{abstract}

\end{frontmatter}


\section{Introduction}
Transferring a symbolic \emph{Mathematica} expression to \emph{Matlab} is a recurring issue one for anyone using these two programs. On the one hand, the \emph{Mathematica} software is perfect to work out analytical calculations and derive involved symbolic formulas. On the other hand,  \emph{Matlab} offers a quasi-infinite and very flexible environment for graphical, numerical or statistical treatment. The perspective of working in both environment to solve a given problem is thus very attractive. As long as the equations involved remain simple, an expression derived in \emph{Mathematica} can be straightforward reproduced in \emph{Matlab}. Problems arise for long and intricate formulas which cannot be easily copy-pasted. The transfer of a polynomial poses an additional problem: if one is willing to take advantage of the \emph{Matlab} polynomial routines, the very definition of the function needs to be adapted because \emph{Matlab} defines a polynomial from its coefficients.

The \emph{Mathematica} Notebook presented in this paper offers a simple solution to this problem. Given any polynomial expression derive in \emph{Mathematica}, the Notebook generates a text which can be directly copy-pasted into \emph{Matlab}.

\section{\label{sec:notebook}Notebook description}
 Consider a polynomial of the form,
 \begin{equation}\label{eq:poly}
    P(x)=\sum_{i=0}^n a_i(\alpha_1\ldots \alpha_k)x^i,
 \end{equation}
where the coefficients $a_{1\ldots n}$ are themselves polynomial expressions of some parameters $\alpha_{1\ldots k}$. While both \emph{Mathematica} and \emph{Matlab} can deal with this kind of expressions, Matlab requires a different format if functions such as $roots$ are to be exploited. More specifically, the polynomial $P$ defined by Eq. (\ref{eq:poly}) will be expressed in \emph{Matlab} under the form of a 1D matrix,

\textbf{p = [$a_n(\alpha_1\ldots \alpha_k)$ ~ $a_{n-1}(\alpha_1\ldots \alpha_k)$ ~ \ldots ~ $a_0(\alpha_1\ldots \alpha_k)$]}

The Notebook which is about to be described generates the format needed by \emph{Matlab} from the \emph{Mathematica} expression. The starting point is the definition of the polynomial,

\emph{In[1]:=}\textbf{P =$a_n(\alpha_1\ldots \alpha_k)x^n+a_{n-1}(\alpha_1\ldots \alpha_k)x^{n-1}$ + \ldots + $a_0(\alpha_1\ldots \alpha_k)$;}

Note that the polynomial does not need to be specifically formatted like in Eq. (\ref{eq:poly}). The Notebook can perfectly deal with an expression where the coefficients of the $x^i$'s are not factorized. If the polynomial expression accounts for Greek characters (dealt with by \emph{Mathematica}), they will have to be replaced by some equivalents for proper \emph{Matlab} treatment. For example, if the \emph{Mathematica} polynomial accounts for the parameters $\beta,\gamma$ and $\omega$, the next line of the Notebook is,

\emph{In[2]:=}\textbf{PClear = P/. $\{\beta \rightarrow$ beta, $\gamma\rightarrow$ gamma, $\omega\rightarrow$ omega\};}

We then need the degree of the Polynomial, given by

\emph{In[3]:=}\textbf{Deg = Exponent[PClear, x];}

We now have \emph{Mathematica} extracting the $a_i$'s with,

\emph{In[4]:=}\textbf{Coefs = Table[InputForm[FullSimplify[Coefficient[PClear, x, k]]], \{k, 0, Deg, 1\}];}

Here, the $k^{\mathrm{th}}$ coefficient is first gathered through the \textbf{Coefficient[PClear, x, k]} command. The resulting expression is simplified with \textbf{FullSimplify}. Note that a simple \textbf{Simplify} can be inserted here if the argument is too involved for \textbf{FullSimplify} to execute successfully.

The \textbf{InputForm} command is a key element of the transfer to \textbf{Matlab}. \emph{Mathematica} can deal with symbolic fractions or power, but not \emph{Matlab}. For example, the \emph{Mathematica} expression,

   \begin{equation}
    \frac{a^2b^3}{c^4(t-u)}\nonumber
   \end{equation}

must be cast under the form

a\^{}2*b\^{}3/(c\^{}4*(t-u)),

for \emph{Matlab} to interpret it. The \textbf{InputForm} command is responsible for the \emph{Matlab} formatting of the polynomial coefficients. Finally, the \textbf{Table} command ensures that the operation is implemented for all the $a_i$'s, and the result is stored in the \emph{Mathematica} array \textbf{Coefs}.

We finally have Mathematica displaying the content of the array \textbf{Coefs} under a form which can be copy-pasted to \textbf{Matlab},

\emph{In[2]:=}\textbf{Do[Print[``P('', Deg + 2 - n, ``)='', Coefs[[n]], ``;''], \{n, Deg + 1\}];}

\begin{center}
\begin{figure}
\includegraphics[width= \textwidth]{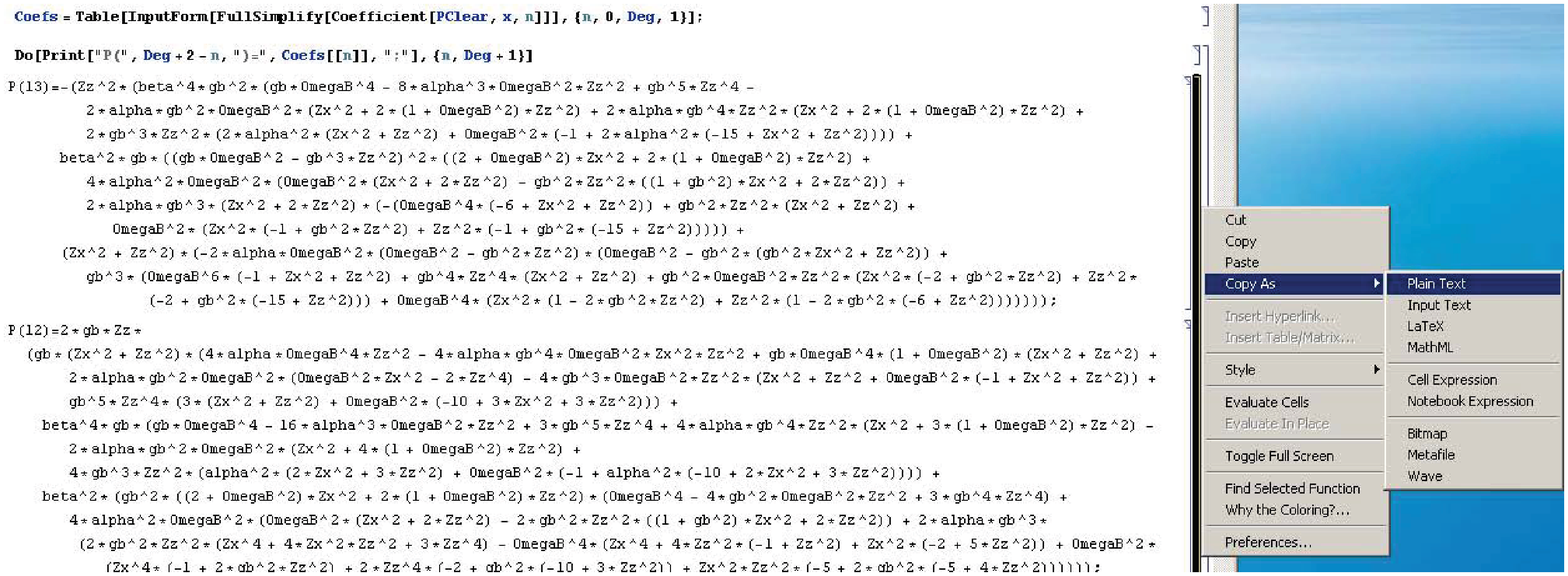}
\caption{Copy-pasting the \emph{Mathematica} output to \emph{Matlab}.}
\label{fig}
\end{figure}
\end{center}

The command above thus prints all the coefficients of the polynomial from P(1), coefficient of $x^n$ to P(deg+1), coefficient of $x^0$. The output can then be copy-pasted to a ``.m'' \emph{Matlab} file. Note that it is here necessary to ``Copy As Plain Text'' from \emph{Mathematica}, as illustrated in Figure \ref{fig}.

\section{Conclusion - An example}
Two files have been attached to this article, as an illustration of the kind of transfer allowed by this technique. The \emph{Mathematica} Notebook ``Bridge to MatMab.nb'' introduces a imposing polynomial which arises as the dispersion equation of a quite involved relativistic magnetized beam plasma system \cite{Bret}. It is a very involved polynomial in terms of the parameters $x,Zx,Zz,\alpha,\gamma_b,\beta,\Omega_B$. The degree of the polynomial in terms of $x$ is 14. The expression results from the determinant of a dielectric tensor, and the coefficients of $x$ are not gathered at all. The execution takes only a few minutes, which can be reduced even further by replacing the \textbf{FullSimplify} by the faster \textbf{Simplify} command in the coefficients calculation.

After copying the output as shown on Figure \ref{fig}, the result has been directly pasted in the \emph{Matlab} file ``poly.m'' and can be readily exploited in this Software.

\section{Acknowledgements}
This work has been  achieved under projects ENE2009-09276 of the
Spanish Ministerio de Educaci\'{o}n y Ciencia and PAI-05-045 of
the Consejer\'{i}a de Educaci\'{o}n y Ciencia de la Junta de
Comunidades de Castilla-La Mancha.

\end{document}